\documentclass[twocolumn,superscriptaddress,showpacs,preprintnumbers,amsmath,amssymb]{revtex4}

\usepackage{amsfonts}
\usepackage{amssymb}
\usepackage{graphicx}
\usepackage{color}

\def\bce{\begin{center}}
\def\ece{\end{center}}
\def\be{\begin{equation}}
\def\ee{\end{equation}}
\def\bea{\begin{eqnarray}}
\def\eea{\end{eqnarray}}

\newcommand{\bs}{\begin{subequations}}
\newcommand{\es}{\end{subequations} \noindent}
\newcommand{\ba}{\begin{array}}
\newcommand{\ea}{\end{array}}
\newcommand{\bi}{\begin{itemize}}
\newcommand{\ei}{\end{itemize}}

\newcommand{\bac}{\begin{array}{c}}

\begin{document}


\title{Symmetry breaking in the collisions of double channel BEC solitons}

\author{Nguyen Viet Hung, Pawel Zi\'n, Eryk Infeld}

\address{National Centre for Nuclear Research, ul. Ho\.{z}a 69, PL-00-681 Warsaw, Poland}

\author{Marek Trippenbach\footnote{Corresponding author: e-mail: matri@fuw.edu.pl}}

\address{Institute of Theoretical Physics, University of Warsaw, ul. Ho\.{z}a 69, PL--00--681 Warszawa, Poland. }

\begin{abstract}
We investigate an attractive Bose-Einstein condensate in two coupled one dimensional channels. In this system a stable double channel soliton can be formed. It is symmetric for small interaction parameters and asymmetric for large ones. We study this symmetry breaking phenomenon in detail. Next, we investigate the dynamics of symmetric double channel soliton collisions.  For sufficiently strong interactions we observe spontaneous symmetry breaking during the collision. Approximate considerations based on two different methods, Bogoliubov and variational, are used to describe this effect. The results are compatible.
\end{abstract}

\maketitle

\section{Introduction}

Directional couplers have been studied extensively in the context of all-optical soliton switching after the pioneering work of Jensen \cite{Jensen} and Trillo {\it et al} \cite{Trillo}. These ideas were developed and applied in  fiber-optic devices which require splitting of an optical field into two coherent but physically separate parts.  Optical fiber couplers have been studied for their potential applications to ultra fast all optical switching processing, such as an optical
switch \cite{Ramagnoli,Uzunov,Agrawal,Trillo2,Ramos,Kumar,Sarma,Wang,Skiner}. Numerous studies, including soliton switching in dual-core optical fibers have shown excellent switching
characteristics, with efficiencies around 96 \% for a wide range of input energies \cite{Umarov,Malomed,EI1,EI2,Belanger,Abdullaev,Malomed2,Malomed3,Finlayson,Chu}.
A review of the basic ideas and the literature  can be found in Saleh \cite{Saleh}. Recently, nonlinear directional couplers with dissimilar cores have attracted
attention, as several new effects can occur in them \cite{Kaup1,Kaup2,Lakoba}.The study of nonliear couplers is no longer confined to the conventional silica based optical fiber coupler. It has recently been extended to AlGaAs nanowire \cite{Malomed} and lead silicate based holey fiber couplers \cite{Sarma2}.

The order of the paper is as follows. First we introduce our one dimensional model and next find a solution in the form of pairs of solitonic wavepackets, one in each channel. We use the variational approximation and the sechans shaped ansatz. We plot a bifurcation diagram showing that the symmetric variational states become unstable for large enough $N$. In the next section we investigate two soliton pair collisions. Here we introduce the Bogolibov analysis and, in the following section the variational approximation to describe modulational instability caused by the collision. Comparison of the two methods winds up the paper. The results are not identical, but are compatible.

\section{The model}

\begin{figure}[tbp]
\includegraphics[width=8.5cm]{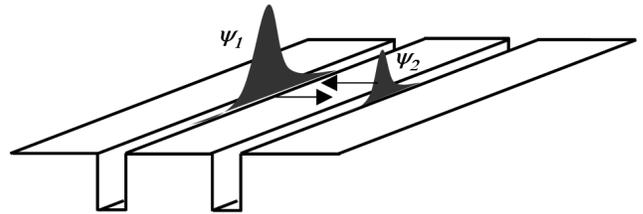}
\caption{Schematic view of the single soliton solution in the double channel potential.} \label{fig1}
\end{figure}

Our first system consists of an attractive Bose Einstein condensate in the potential of two quasi one dimensional channels and is schematically shown in Fig.(\ref{fig1}). This model can also be used to describe light propagation in  coupled nonlinear fibers. Here we refer to the BEC case; this merely determines the range of parameters (physical coefficients) appearing in the equations. The dynamics are governed by two coupled nonlinear Schrodinger equations (NLS):
\begin{eqnarray} \nonumber
i\hbar \partial_t \psi_{1}(x,t)  = -\frac{\hbar^2}{2m}\partial_x^2 \psi_{1}(x,t)  &-& g|\psi_{1}(x,t)|^{2}\psi_{1}(x,t)
\\
&-& \kappa\psi_{2}(x,t) \label{modelo} \nonumber
\\ \\
i\hbar \partial_t \psi_{2}(x,t) = -\frac{\hbar^2}{2m} \partial_x^2 \psi_{2}(x,t)  &-& g|\psi_{2}(x,t)|^{2}\psi_{2}(x,t) \nonumber
\\ \nonumber
&-& \kappa\psi_{1}(x,t), \nonumber
\end{eqnarray}
with the total number of particles in the condensate equal to
\begin{equation} \label{eq:norm}
N_{p}=\int \mbox{d} x \, \left( |\psi_{1}(x,t)|^2 +  |\psi_{2}(x,t)|^2 \right).
\end{equation}
Here we use an approximation consisting of linear coupling between effectively one dimensional channels
Simple rescalling of the system parameters, assuming $\kappa > 0$, leads to the set of reduced equations
\begin{eqnarray}
i \partial_t \psi_{1} &=& -\frac{1}{2} \partial_x^2 \psi_{1} -|\psi_{1}|^{2}\psi_{1} -\psi_{2}\nonumber
\\ \label{model1}
\\
i \partial_t \psi_{2} &=& -\frac{1}{2} \partial_x^2 \psi_{2}-|\psi_{2}|^{2}\psi_{2}-\psi_{1}\nonumber
\end{eqnarray}
with  total norm $N$ as in Eq.~(\ref{eq:norm}).

The relation between the norm $N$ and the number of particles $N_{p}$ is given by  $N=gN_{p}\sqrt{m/(\hbar^2\kappa)}$. Notice that after  rescalling, the system is fully characterized by {\it just one parameter} $N$. It has a known one soliton solution, as was shown in \cite{Progress2002}. In the next section we present a short summary of one soliton solutions and analyze their stability.

\section{One soliton solutions} \label{Sec2}

One of the methods commonly applied in the quest for stationary states (solitons) is the variational approximation. A detailed account of this technique in the context of solitons can be found, for instance, in Ref.~\cite{Progress2002}. In this approach we first identify the energy $E$ of our system:
\begin{equation}
E=\int \mbox{d} x \left (\frac{1}{2}\sum_{j=1}^{2} \left|\partial_x \psi_j \right|^2 - \frac{1}{2}\sum_{j=1}^{2}|\psi_j|^4 -  \left
(\psi_1\psi_2^* + c.c.\right ) \right )
\end{equation}
Next we introduce the variational Ansatz for the modulus of $\psi_{i}(x)$
\begin{eqnarray}
|\psi_{1,2}(x)|=\sqrt{\frac{N\left(1 \pm
z\right)}{4W}} \ \textmd{sech}
\left(\frac{x}{W}\right), \label{Ansatz}
\end{eqnarray}
where $W$ is the width of the soliton and $z$ is the asymmetry parameter defined as:
\begin{equation}\label{z}
z = \frac{1}{N} \int \mbox{d} x \, \left( |\psi_{1}|^2 -  |\psi_{2}|^2  \right).
\end{equation}
Note that when $z=0$ and $W=4/N$ this can represent an exact solution. It is the best possible trial function as we will see.

\begin{figure}[tbp]
\includegraphics[width=9.5cm]{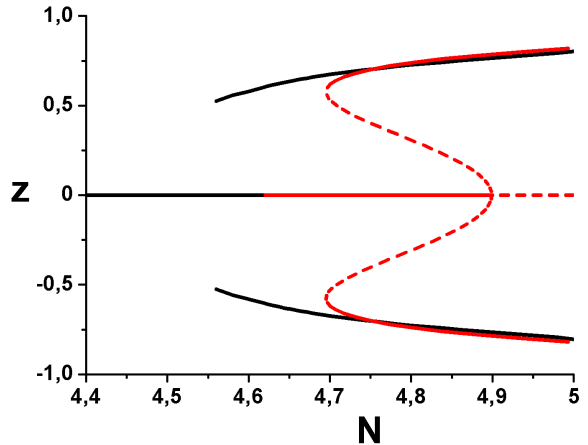}
\caption{Diagram of stable solutions. The thin solid lines represent the variational stationary stable states whereas the dashed ones represent the variational unstable stationary states.
The stable states obtained in a direct numerical simulation are marked by thick lines.} \label{fig2}
\end{figure}

Within the class of trial-functions (\ref{Ansatz}) the energy can be expressed in terms of variational parameters upon performing integration over $x$:
\begin{equation}\label{Ham}
E = \frac{N}{6W^2} -   \frac{N^2(1+z^2)}{12W} - N\sqrt{1-z^2}.
\end{equation}
Notice that $E$ now depends on the norm $N$ and variational parameters $W$ and $z$. With increasing value of $z$, the interaction energy $-\frac{N^2(1+z^2)}{12W}$ decreases, while the tunneling energy $- N\sqrt{1-z^2} $ increases. Moreover, the interaction energy scales like $N^2$, while the tunneling energy scales like $N$. Hence, for small values of $N$ stable states will be symmetric ($z=0$), while for $N$ above a certain critical value they become asymmetric. This is summarized in Fig.(\ref{fig2}). Stationary solutions can be found considering $\partial E/\partial W=\partial E/\partial z=0$, which leads to our friend $W=4/N$ as in the exact solution, and next to the condition
\begin{equation}
z\left (\frac{24}{N^2}-(1+z^2)\sqrt{1-z^2}\right )=0.
\end{equation}
This equation is satisfied in two cases: $z=0$ (symmetric case) or  $z^6+z^4-z^2+\left(\frac{24}{N^2}\right)^2-1=0$ (asymmetric). The latter is a cubic for $z^2$. To find which states are stable, we compute the second order derivatives of the reduced energy (\ref{Ham}). The results are presented in Fig.(\ref{fig2}), where stable solutions are denoted by solid lines and unstable ones by dashed lines. For comparison we also included the stable states obtained from direct numerical computation of the NLS equation. This Figure provides clear evidence that in our case we have a subcritical transition with hysteresis.

The variational approximation correctly predicts the shape of the hysteresis, but is unfortunately not very accurate as to its position. Our calculations show that the norm of symmetric solution when it looses its stability is equal to $N_{cr,v}=\sqrt{24}\approx 4.90$ in the variational approximation, whereas in direct numerical simulation we obtain the value $N_{cr}\approx 4.62$). This is not very surprising since the value of $N_{cr,v}$ depends on the variational Ansatz. Departure from Eq.~(\ref{Ansatz}) leads to less accurate values. For example, a different Ansatz proportional to $\exp[(-x/W)^2]$, yields for $N_{cr,v}=\sqrt{8\pi} \sim \sqrt{25.12} \sim 5.01$.

Notice that our localized solutions in the symmetric case reduce to ordinary soliton solutions for the NLS equation. If we admit an asymmetry, the system we consider here is not integrable. In what follows we will call localized double channel solutions as considered above double channel solitons (DCS)

\section{Two soliton collisions}

\begin{figure}[tbp]
\includegraphics[width=8.5cm]{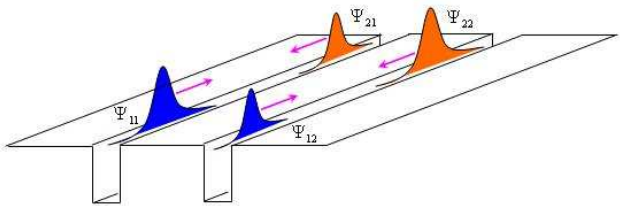}
\caption{Schematic of a collision of two double channel solitons.} \label{schemat_colis}
\end{figure}

In this section we study the collisions of pairs of identical DCS's, each of which has  norm equal to $N$. This is schematically illustrated in Fig.~(\ref{schemat_colis}). We expect that, when the total norm exceeds the critical value discussed above ($2N>N_{cr}$) the system becomes unstable during the collision, especially when the overlap of the wavefunctions is substantial. The main subject of the present section is the study of instability in the Bogolibov manner. For this purpose we add small initial asymmetric perturbations to both DCS. These perturbations will be amplified during the collision due to the instability. We analyze the growth of the instability by monitoring the asymmetry parameter in time
\begin{equation}\label{AP}
z(t) = \frac{1}{2N}\int_{-\infty}^{\infty}\left(|\psi_{1}(x,t)|^2-|\psi_{2}(x,t)|^2 \right)\textmd{d}x.
\end{equation}

\begin{figure}[tbp]
\includegraphics[width=8.5cm]{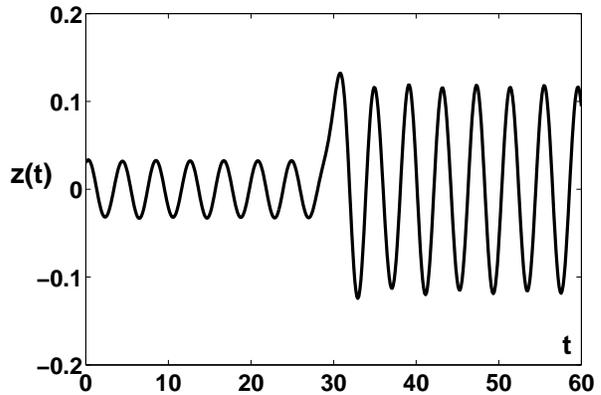}
\caption{The asymmetry parameter versus time. We observe an amplification of the oscillations during the collision due to instability.} \label{figur}
\end{figure}

An example of the collisional event is illustrated in Fig.(\ref{figur}). Here we plot the asymmetry parameter versus time. At the initial time, when two DCS's are far apart we add a small asymmetric perturbation. This perturbation causes the asymmetry parameter to oscillate. Asymptotically, before and after the collision,oscillations are harmonic, with constant amplitude. Interaction during the collision can amplify the perturbation, as seen in the Figure (\ref{figur}). In the regime of small perturbations the final amplitude of the oscillation is directly proportional to the initial amplitude of oscillation (the Bogoliubov approximation is valid). Hence  we choose the ratio of the final amplitude to the initial one as a measure of instability and label it the amplification ($R$). To identify the parameters that affect the amplification, we apply the Bogoliubov approximation.

\subsection{Bogoliubov analysis}

In this section we introduce the Bogoliubov method and tailor it to our problem. We add a small perturbation to the symmetric solution
\begin{eqnarray} \nonumber
\psi_1(x,t) &=& \psi(x,t) + \delta_1(x,t),
\\ \nonumber
\psi_2(x,t) &=& \psi(x,t)+\delta_2(x,t),
\end{eqnarray}
and linearize the two NLS equations, obtaining a set of Bogolubov equations for the $\delta_i$ functions:
\begin{eqnarray}
i\partial_t \delta_{1} &=&
-\frac{1}{2} \partial_x^2\delta_{1} -2|\psi|^{2}\delta_{1}-\psi^2\delta_1^{*}-\delta_{2}, \nonumber
\\ \label{Bogolub:Eq}\\
i\partial_t \delta_{2} &=&
-\frac{1}{2}\partial_x^2 \delta_{2} -2|\psi|^{2}\delta_{2}-\psi^2\delta_2^{*}-\delta_{1}. \nonumber
\end{eqnarray}
Notice that we restrict our considerations to the initially symmetric case. Hence $\psi(x,t)$, which can be treated as a core of the wavefunction, the same in each channel. It represents the two soliton collision in 1D NLS:
\begin{equation} \label{gp1}
i \partial_t \psi = -\frac{1}{2} \partial_x^2 \psi -|\psi|^{2}\psi-\psi,
\end{equation}
normalized
\begin{equation}\label{fNorm}
\int_{-\infty}^{\infty}|\psi(x,t)|^2\textmd{d}x=N.
\end{equation}
As the function $\psi$ appears in both channels the total norm of the unperturbed solution is $2N$.

The two soliton solution of Eq.~(\ref{gp1}) is known \cite{ZaiDong}.
Far away from the collisional region we obtain, as a special, symmetric case for large $d$
\begin{eqnarray}
\psi(x,t) &=& \frac{N}{4} \textmd{sech} \left(\frac{N}{4} \left(x-d + vt \right)\right)e^{i(\mu t - \frac{v^2}{2} t +  v (x-d) + \theta/2 )} \nonumber
\\ \label{2solit}
&+& \frac{N}{4} \textmd{sech} \left(\frac{N}{4} \left(x + d  -vt\right)\right)e^{i(\mu t -  \frac{v^2}{2} t -  v (x+d) - \theta/2)},
\end{eqnarray}
where $\mu = 1 + N^2/32$, the distance between the maxima is $2d$ and $\theta$ is the phase difference between the solitons.

To decouple the Bogoliubov equations (\ref{Bogolub:Eq}), we define symmetric and antisymmetric combinations of the deltas
\begin{eqnarray}
\delta_{S,D}(x,t) &=& \delta_1(x,t) \pm \delta_2(x,t), \nonumber
\end{eqnarray}
and obtain
\begin{eqnarray}
i \partial_t \delta_{S} &=& -\frac{1}{2} \partial_x^2 \delta_{S}
-2|\psi|^{2}\delta_{S}-\psi^2\delta_S^{*}-\delta_{S},\label{sum}
\\ \nonumber \\
i\partial _t \delta_{D} &=& -\frac{1}{2} \partial_x^2 \delta_{D}
-2|\psi|^{2}\delta_{D}-\psi^2\delta_D^{*}+\delta_{D}, \label{diff}
\end{eqnarray}
where $\psi$ is given by Eq.~(\ref{2solit}). When the two identical symmetric DCS's are well separated (long before and long after the collision) they are practically independent. The initial state $\psi(x,0)$, which represents two solitons ready to collide, is uniquely defined by three parameters: the norm $N$, $v$ and $\theta$. We assume that $d$ is large and fixed. In the following we restrict perturbations to a specific class. We take a two soliton solution $\psi(x,0)$ as above and define perturbed wavefunctions in both channels as
\begin{eqnarray} \nonumber
\psi_1(x,0) &=& \sqrt{1+z_0} \psi(x,0) \exp (i \varphi_0/2)
\\ \label{pert1}
\\ \nonumber
\psi_2(x,0) &=& \sqrt{1-z_0} \psi(x,0) \exp ( -i \varphi_0/2).
\end{eqnarray}
The advantage of this parametrization is that it matches the condition $z_0=z(0)$, where $z(0)$ is defined in Eq.~(\ref{AP}).
Next we linearize wavefunctions (\ref{pert1}) in perturbation parameters $z_0$ and $\varphi_0$ obtaining
\begin{eqnarray}
\delta_S(x,0) &=& 0
\\
\delta_D(x,0) &=& (z_0 - i \varphi_0)\psi(x,0)
\end{eqnarray}
Finally we introduce perturbation variables $\alpha$ and $\vartheta$ defined as
\begin{equation} \label{rel}
z_0 - i \varphi_0 = \alpha \exp(i \vartheta) \Longrightarrow \delta_D(x,0) =  \alpha \exp(i \vartheta)\psi(x,0).
\end{equation}
In the Bogoliubov method the asymmetry parameter (\ref{AP}) is given by:
\begin{equation}\label{APB}
z(t) = \frac{1}{2N} \int \mbox{d} x \ ( \psi \delta_D^* + \psi^* \delta_D ).
\end{equation}
It is linear in $\delta_D$ and the amplification $R$ does not depend on $\alpha$. Hence in our numerical studies presented below we consider $R$ as a function of four parameters: $N$, $v$,  $\theta$ and $\vartheta$. Before referring to the numerical results we show that some insight can be obtained using variational approximation.

\subsection{Variational approximation}

 Consider as our Ansatz
\begin{eqnarray}
\psi_{1}(x,t)&=&\sqrt{1+z(t)}\psi(x,t)\textmd{exp}\left(\frac{i\varphi(t)}{2}+i\phi(t)\right),
\nonumber \\ \label{variational}
\\
\psi_{2}(x,t)&=&\sqrt{1-z(t)}\psi(x,t)\textmd{exp}\left(-\frac{i\varphi(t)}{2}+i\phi(t)\right),\nonumber
\end{eqnarray}
where $\psi(x,t)$ is again a two soliton solution of equation (\ref{gp1}).
In this approach we recognize three variational functions $z(t)$, $\phi(t)$ and $\varphi(t)$. The new phases $\phi$ and $\varphi$ replace $\theta$ and $\vartheta$. Clearly $z(t)$ is equal to the asymmetry parameter defined in (\ref{AP}). Analogously the parameter $\varphi(t)$ is introduced to match $\varphi_0$ defined by (\ref{pert1}) at $t=0$.

The Lagrangian corresponding to equation (\ref{model1}) is
\begin{eqnarray} \label{Lagreduce}
L&=&\int_{-\infty}^\infty \left
[\frac{1}{2}\sum_{j=1}^{2}i\left(\psi_j\partial_t\psi_j^{*}-\psi^{*}_j\partial_t\psi_j\right)-\frac{1}{2}\sum_{j=1}^{2}|\partial_x\psi_j|^2 \right.
\nonumber \\  && \left.~~~~~~~~
+\frac{1}{2}\sum_{j=1}^{2}|\psi_j|^4+\left
(\psi_1\psi_2^*+\psi_2\psi_1^*\right ) \right ]\textmd{d}x,
\end{eqnarray}
and after substituting the Ansatz (\ref{variational})and using (\ref{fNorm}) we reduce it to the following form
\begin{eqnarray}\label{Lagefredu2}
L&=&2N\cos\varphi\sqrt{1-z^2}+N(2\dot{\phi}+z\dot{\varphi})\nonumber \\ \nonumber \\&+&(z^2+1)g(t) - 2h(t),
\end{eqnarray}
where
\begin{eqnarray} \label{g}
g(t)&=&\int_{-\infty}^{\infty}|\psi(x,t)|^4\textmd{d}x,
\\ \nonumber \\
h(t)&=&\int_{-\infty}^{\infty}|\partial_x\psi(x,t)|^2\textmd{d}x. \nonumber
\end{eqnarray}
The Euler-Lagrange equations for variational functions $ z(t)$ and $\varphi(t)$ read
\begin{eqnarray}\label{VE}
\dot{\varphi}+z\left(\frac{2g(t)}{N}-\frac{2\cos\varphi}{\sqrt{1-z^2}}\right) &=& 0, \nonumber
\\ \\
\dot{z}+2\sqrt{1-z^2}\sin\varphi &=& 0. \nonumber
\end{eqnarray}
The equation for $\phi$ is not included here (it just represents norm conservation).
To be able to make a comparison with Bogoliubov analysis (the limit $|z(t)| \ll 1$ and $|\varphi(t)| \ll 1$) we linearize the above equations
\begin{eqnarray}\label{Lagefredu4}
\dot{\varphi} &=& 2z\left(1-\frac{g(t)}{N}\right), \nonumber
\\ \\
\dot{z}&=&-2\varphi \nonumber
\end{eqnarray}
and combine them to obtain an equation for $z(t)$
\begin{equation}\label{insta1}
\ddot{z}+4z\left(1-\frac{g(t)}{N}\right)=0.
\end{equation}
To understand equation (\ref{insta1}) we use the analogy of a classical ball on a spring. If we denote the position of the ball by $z$ and its velocity by $\varphi$, the last term in equation (\ref{insta1}) represent a harmonic force with time dependent spring constant $4\left(1-\frac{g(t)}{N}\right)=\omega^2$, see Fig.~(\ref{figur}). The function $g(t)$ depends on the overlap of the colliding solitons. Long before and long after the collision, when solitons are practically independent, the coefficient $g(t)$ is constant and equal to $\frac{N^3}{24}$. Therefore asymptotically the spring constant is equal to $\left(1-\frac{N^2}{24}\right)$. If we refer to previous considerations in section \ref{Sec2} we recognize that it can be rewritten as $\left(1-\frac{N^2}{N_{cr,v}^2}\right)$, where $N_{v,cr}$ is the critical norm given by the variational approximation. For that reason the spring constant is asymptotically positive when the single DCS is stable and $z(t)$ performs harmonic oscillations before and after the collision, as seen in Figure (\ref{figur}).

The dynamics of the collision can be interpreted in terms of the ball on spring analogy. During the collision the value of the spring constant can be negative. Then the harmonic force acting on our ball changes its character from attractive to repulsive. It can add energy to the system increasing the amplitude of oscillations, which is the case shown in fig.~(\ref{figur}). But it can conversely decrease the amplitude when the ball is moving towards the center. As we see the amplification depends crucially on the initial velocity $\varphi(0)=\varphi_0$. In conclusion the amplification process is very sensitive to the phase parameter $\varphi$.

\subsection{Comparison of the two methods}

\begin{figure}[tbp]
\begin{center}
\includegraphics[width=8.5cm]{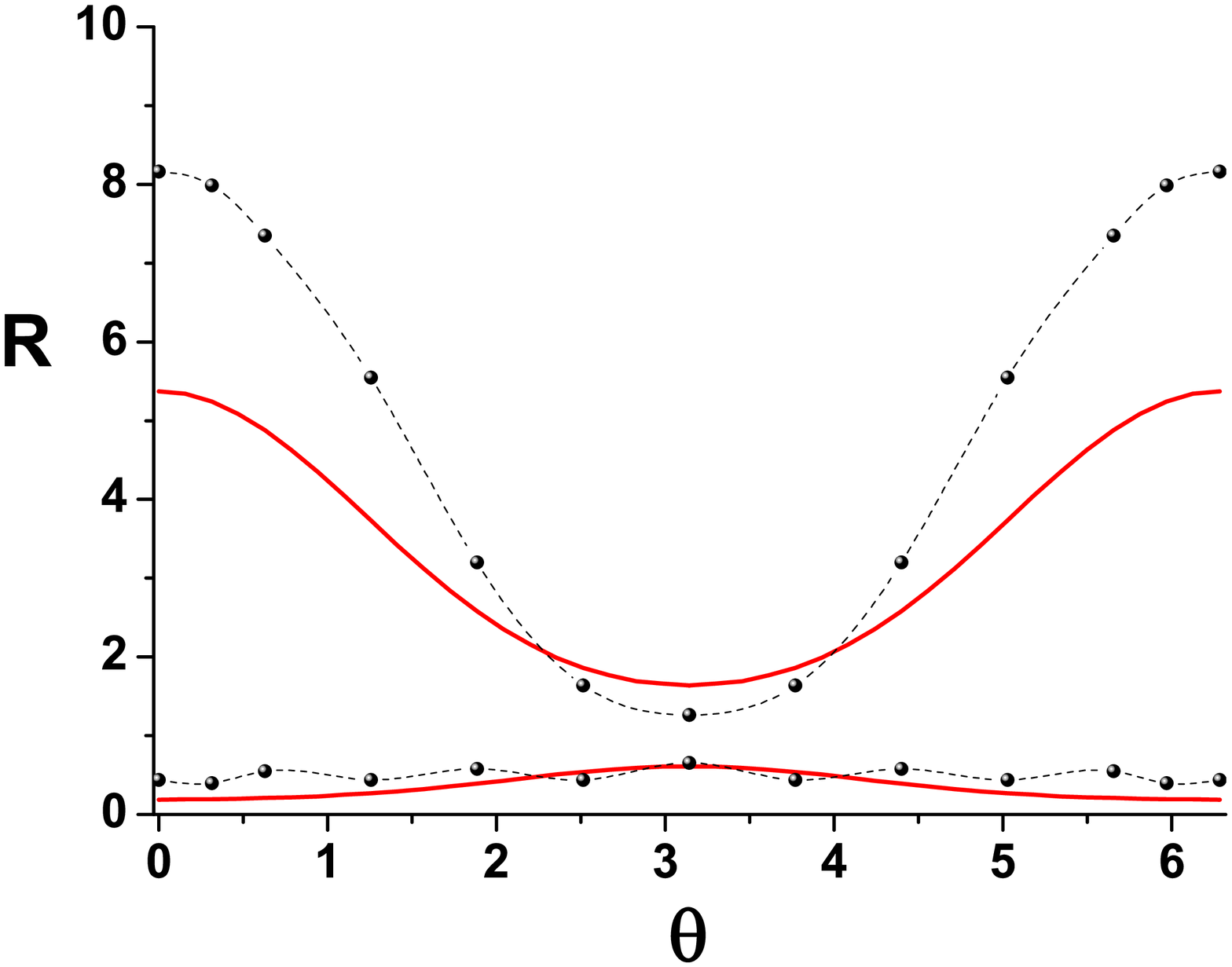}
\includegraphics[width=8.5cm]{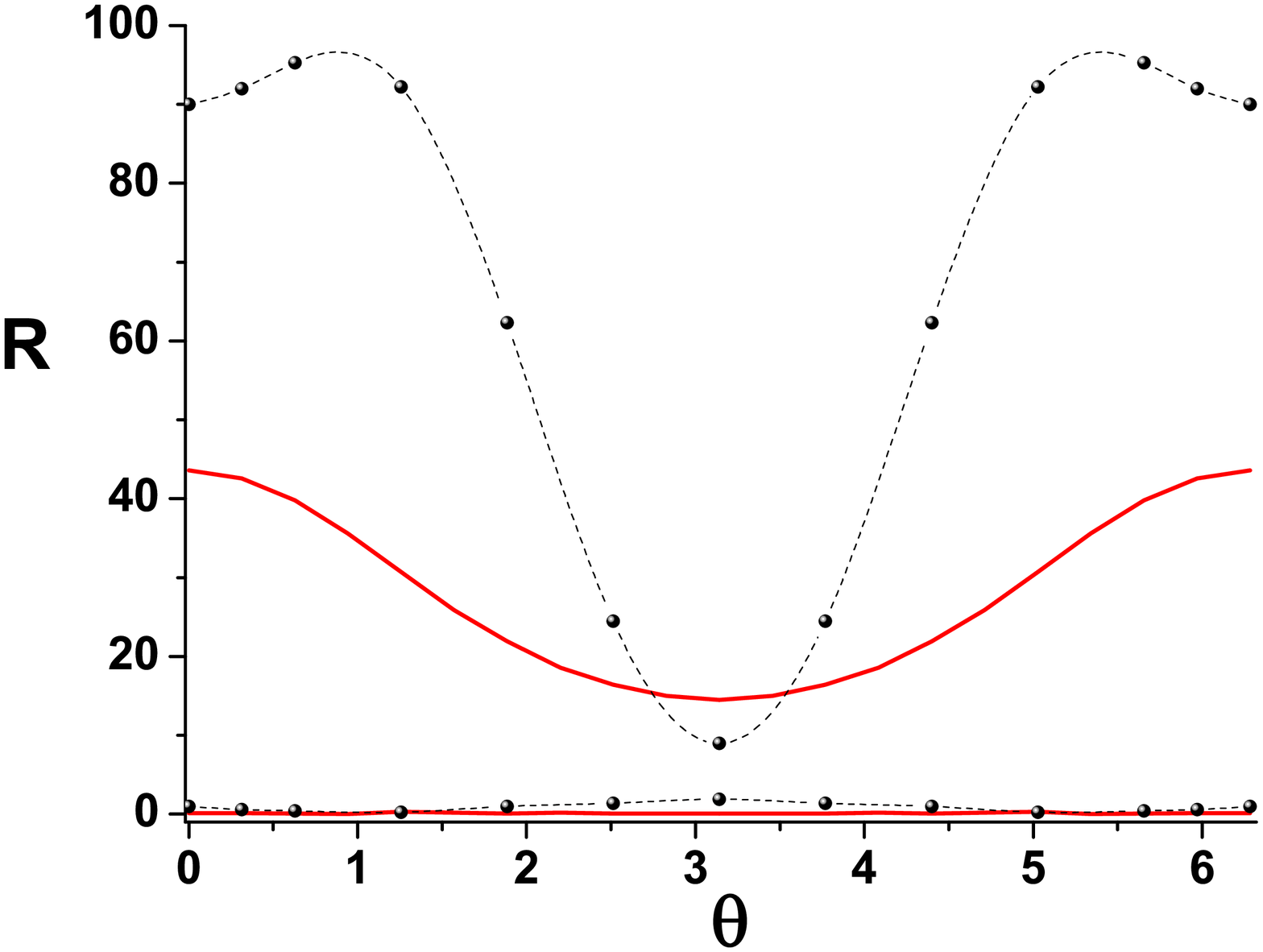}
\caption{Amplification $R$ versus relative phase $\theta$ between solitons. The points show the results of the Bogolubov method while the lines illustrate variational approximation predictions. We observe symmetric structures around the center $\theta_0=\pi$. The top panel corresponds to the case $N=3$, $v=0.75$ when amplification is small. The bottom panel show large amplification in the case $N=4$ and $v=0.5$. In both cases the minimal curves are the better fit} \label{fig30}
\end{center}
\end{figure}

In this section we present the results of a comparison between the variational approximation and the Bogoliubov method. We choose the initial values $z(0)$ and $\varphi(0)$ in the variational Ansatz and initial values $z_0$ and $\varphi_0$ in Eq.~(\ref{pert1}) in the numerical calculations. Additionally we need to rescale the function $g(t)$ since the value of the critical norm $N_{cr}$ is slightly larger that the variational estimate $N_{cr,v}$. In our case the rescaling factor is $N_{cr}/N_{cr,v} =1.124$.

\begin{figure}[tbp]
 \includegraphics[width=8.5cm]{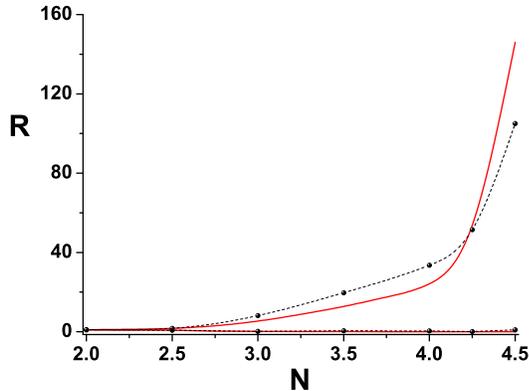}
\caption{Amplification $R$ versus total norm $N$. The points follow from the Bogolubov method as described in the text, while the continuous lines were obtained by the variational approximation. We observe the threshold of amplification around $N_{cr}/2$. The Bogolibov results were obtained by varying $\Theta$ and $\vartheta$.}  \label{fig40}
\end{figure}

The results of our studies are summarized in figures (\ref{fig30}) and (\ref{fig40}). To obtain the first figure we fixed the values of $N$ and $v$. Next for each value of relative phase $\theta$ we found the range of amplification $R$ corresponding to the full range of $\vartheta$. We chose the minimal and maximal values of $R$ and marked them on the figure. We show the results for two particular cases. First we took a soliton norm to be slightly above the threshold $N_{cr}/2$, and observed small amplification. Then we increased the norm and decreased the velocity to obtain much larger amplification. In both cases we observed qualitative agreement between numerical results and the variational approximation. We also notice that the range of amplification, corresponding to different values of $\vartheta$ is very wide.

The main result is presented in Fig.~(\ref{fig40}). In this study we fix the velocity of the solitons and vary their norm. For each value of the norm we find the minimum and maximum of the amplification with respect to different values of both $\theta$ and $\vartheta$. We see that the maximal amplification becomes significant when the norm exceeds $N_{cr}/2$.

\subsection{Results and conclusions}

We have been able to apply two different models to double channel BEC soliton collision. These are the variational and, less popular Bogoliubov model. Fortunately, results are similar and the interactions are well described by both models. The main physical effect so described is the enhancement (or suppression) of the amplitude of small oscillations on two colliding pairs of solitons in the double channel case. This phenomenon turned out to be extremely sensitive to the relative phase of the colliding partners.

\section{Acknowledgements}

M.T. was supported by the Polish National Science Center.


\begin{thebibliography}{99}


\bibitem{Jensen} S.M. Jensen, IEEE J. Quantum Electron. 18 (1982) 1580.
\bibitem{Trillo} S. Trillo, S. Wabnitz, E.M. Wright, G.I. Stegeman, Opt. Lett. 13 (1988) 672.
\bibitem{Ramagnoli} M. Romagnoli, S. Trillo, S. Wabnitz, Soliton switching in nonlinear coupler, Opt. Quant. Electron. 24 (1992), Sl2337-S1267.
\bibitem{Uzunov} I.M. Uzunov, R. Muschall, M. Golles, Y.S. Kivshar, B.A. Malomed, F. Lederer, Phys. Rev. E 51 (1995) 2527.
\bibitem{Agrawal} G. Agrawal, Nonlinear Fiber Optics, Academic Press, London, 1995.
\bibitem{Trillo2} S. Trillo, S. Wabnitz, Opt. Lett. 16 (1991) 1.
\bibitem{Ramos} P.M. Ramos, C.R. Paiva, IEEE J. Quantum Electron. 35 (1999) 983.
\bibitem{Kumar} A. Kumar, A.K. Sarma, Opt. Commun. 234 (2004) 427.
\bibitem{Sarma} A.K. Sarma, Jpn. J. Appl. Phys. 47 (Pt.1) (2008) 5493.
\bibitem{Wang} Y. Wang, W. Wang, J. Lightwave Technol. 24 (2006) 1041.
\bibitem{Skiner} I.M. Skinner, G.D. Peng, B.A. Malomed, P.L. Chu, Opt. Commun. 113 (1995) 493.
\bibitem{Umarov} B.A. Umarov, F.Kh. Abdullaev, M.K.B. Wahiddin, Opt. Commun. 162 (1999) 340.
\bibitem{Malomed} B.A. Malomed, I.M. Skinner, R.S. Tasgal, Opt. Commun. 139 (1997) 247.
\bibitem{EI1} P. Zin, E. Infeld, M. Matuszewski, M. Trippenbach, Phys. Rev. A 73, (2006) 022105.
\bibitem{EI2} E. Infeld, P. Zin, J. Gocalek, M. Trippenbach, Phys. Rev. E 74, (2006) 026610.
\bibitem{Belanger} P. Belanger, C. Pare, Phys. Rev. A 41 (1990) 5254.
\bibitem{Abdullaev} F. Abdullaev, R. Abrarov, S. Darmanyan, Opt. Lett. 14 (1989) 131..
\bibitem{Malomed2} B.A. Malomed, Phys. Rev. E 51 (1995) R864.
\bibitem{Malomed3} B.A. Malomed, I.M. Skinner, P.L. Chu, G.D. Peng, Phys. Rev. E 53 (1996) 4084.
\bibitem{Finlayson}  N. Finlayson, W.C. Banyai, E.M. Wright, C.T. Seaton, G.I. Stegeman, T.J. Cullen, C.N. Ironside, Appl. Phys. Lett. 53 (1988) 1144.
\bibitem{Chu}  P.L. Chu, B.A. Malomed, G.D. Peng, J. Opt. Soc. Am. B 10 (1993) 1379.
\bibitem{Saleh} B.E.A. Saleh, M.C. Teich, Fundamentals of Photonics, J. Wiley, New York, 1991.
\bibitem{Kaup1} D.J. Kaup, T.I. Lakoba, B.A. Malomed, J. Opt. Soc. Am. B 14 (1997) 1199.
\bibitem{Kaup2} D.J. Kaup, B.A. Malomed, J. Opt. Soc. Am. B 15 (1998) 2838.
\bibitem{Lakoba} T.I. Lakoba, D.J. Kaup, B.A. Malomed, Phys. Rev. E 55 (1997) 6107.
\bibitem{Sarma2} A.K. Sarma, Jpn. J. Appl. Phys. 47 (Pt.1) (2008) 5493.
\bibitem{Progress2002} B.~A.~Malomed, Progr. Opt. \textbf{43}, 69 (2002).
\bibitem{ZaiDong}  ZaiDong, Ann. Phys. \textbf{322} 2545, (2007).


\end{thebibliography}
\end{document}